\documentstyle[12pt,theorem]{article}
\theoremstyle{plain} \newtheorem{th}{Theorem}[section]
\theoremstyle{plain} \newtheorem{cor}[th]{Corollary}
\theoremstyle{plain} \newtheorem{prop}[th]{Proposition}
\theoremstyle{plain} \newtheorem{lem}[th]{Lemma}
\theoremstyle{plain} \newtheorem{df}[th]{Definition}
{\theorembodyfont{\upshape} \theoremstyle{break}\newtheorem{rem}[th]{Remark}}
\theoremheaderfont{\scshape}
\newcommand{\qed}{\hfill\hbox{\rule[-2pt]{3pt}{6pt}}}
\makeatletter

\newdimen\LENB \newdimen\LENW \newdimen\THI 
\newdimen\LENWH \newdimen\LENTOT \newcount\N 
\def\vbrknlnele#1#2#3{
  \LENB=#1pt \LENW=#2pt \THI=#3pt
  \LENWH=\LENW \divide\LENWH by 2
  \LENTOT=\LENB \advance\LENTOT by \LENW
  \vbox to \LENTOT{
    \vbox to \LENWH{}
    \nointerlineskip
    \vbox to \LENB{\hbox to \THI{\vrule width \THI height \LENB}}
    \nointerlineskip
    \vbox to \LENWH{}
  }}

\def\vbrknln#1{
  \N=#1
  \vcenter{
    \vbox{
      \loop\ifnum\N>0
        \vbox to 4pt{\vbrknlnele{2}{2}{0.1}}
        \nointerlineskip
        \advance\N by -1
      \repeat
  }}}

\def\vbl#1{\hskip-5pt \vbrknln{#1} \hskip-5pt}

\def\hbrknlnele#1#2#3{
  \LENB=#1pt \LENW=#2pt \THI=#3pt
  \LENTOT=\LENB \advance\LENTOT by \LENW
  \vcenter{
    \vbox to \THI{
      \hbox to \LENTOT{
        \hfil
        \vrule width \LENB height \THI
        \hfil}
  }}}

\def\hblele{\hbrknlnele{2}{2.2}{0.1}}

\def\hblfil{\cleaders\hbox{$ \m@th \mkern1mu \hblele \mkern1mu
$}\hfill}

\def\eqnarray{%
\stepcounter{equation}%
\let\@currentlabel=\theequation
\global\@eqnswtrue
\global\@eqcnt\z@
\tabskip\@centering
\let\\=\@eqncr
$$\halign to \displaywidth\bgroup\@eqnsel\hskip\@centering
$\displaystyle\tabskip\z@{##}$&\global\@eqcnt\@ne
\hfil$\displaystyle{{}##{}}$\hfil
&\global\@eqcnt\tw@$\displaystyle\tabskip\z@{##}$\hfil
\tabskip\@centering&\llap{##}\tabskip\z@\cr}

\def\@cite#1#2{\unskip\nobreak\relax
     \def\@tempa{$\m@th{\hbox{[#1]}}$}%
    \futurelet\@tempc\@citexx}

\def\@citexx{\ifx.\let\@tempd=\@citepunct \@tempc\else
    \ifx,\let\@tempd=\@citepunct \@tempc\else
    \let\@tempd=\@tempa\fi\fi\@tempd}
\def\@citepunct{\@tempc\edef\@sf{\spacefactor=\the\spacefactor\relax}\@tempa
    \@sf\@gobble}

\def\citenum#1{{\def\@cite##1##2{##1}\cite{#1}}}
\def\citea#1{\@cite{#1}{}}

\newcount\@tempcntc
\def\@citex[#1]#2{\if@filesw\immediate\write\@auxout{\string\citation{#2}}\fi
  \@tempcnta\z@\@tempcntb\m@ne\def\@citea{}\@cite{\@for\@citeb:=#2\do
    {\@ifundefined
       {b@\@citeb}{\@citeo\@tempcntb\m@ne\@citea\def\@citea{,}{\bf ?}\@warning
       {Citation `\@citeb' on page \thepage \space undefined}}%
    {\setbox\z@\hbox{\global\@tempcntc0\csname b@\@citeb\endcsname\relax}%
     \ifnum\@tempcntc=\z@ \@citeo\@tempcntb\m@ne
       \@citea\def\@citea{,}\hbox{\csname b@\@citeb\endcsname}%
     \else
      \advance\@tempcntb\@ne
      \ifnum\@tempcntb=\@tempcntc
      \else\advance\@tempcntb\m@ne\@citeo
      \@tempcnta\@tempcntc\@tempcntb\@tempcntc\fi\fi}}\@citeo}{#1}}
\def\@citeo{\ifnum\@tempcnta>\@tempcntb\else\@citea\def\@citea{,}%
  \ifnum\@tempcnta=\@tempcntb\the\@tempcnta\else
   {\advance\@tempcnta\@ne\ifnum\@tempcnta=\@tempcntb \else \def\@citea{--}\fi
    \advance\@tempcnta\m@ne\the\@tempcnta\@citea\the\@tempcntb}\fi\fi}

\newcount\N  \newcount\M  \newdimen\SIZE  \newdimen\INC
\def\YGBOX#1#2#3{
      \N=#1  \M=1  \INC=#2pt  \advance\INC by .#3pt  
      \vbox{
         \loop\ifnum\M>0
            \M=\N
            \divide\N by 10         
            \multiply\N by 10
            \advance\M by -\N
            \divide\N by 10
            \SIZE=\INC
            \multiply\SIZE by \M    
            \advance\SIZE by .#3pt
             \hrule  width \SIZE  height .#3pt
              \hbox{\loop\ifnum\M>0                      
                        \vrule  height #2pt  width .#3pt 
                        \hskip #2pt
                        \advance\M by -1  \repeat
                        \vrule  width .#3pt }
             \hrule  width \SIZE  height .#3pt
            \vskip -.#3pt
         \repeat } }

\def\young#1{{             
       \mathchoice{\YGBOX{#1}61}{\YGBOX{#1}61}{\YGBOX{#1}41}{\YGBOX{#1}31}}}
\newcount\N  \newcount\M  \newdimen\SIZE  \newdimen\INC
\def\YGBOXC#1#2#3{
      \N=#1  \M=1  \INC=#2pt  \advance\INC by .#3pt  
      \vcenter{\vbox{
         \loop\ifnum\M>0
            \M=\N
            \divide\N by 10         
            \multiply\N by 10
            \advance\M by -\N
            \divide\N by 10
            \SIZE=\INC
            \multiply\SIZE by \M    
            \advance\SIZE by .#3pt
             \hrule  width \SIZE  height .#3pt
              \hbox{\loop\ifnum\M>0                      
                        \vrule  height #2pt  width .#3pt 
                        \hskip #2pt
                        \advance\M by -1  \repeat
                        \vrule  width .#3pt }
             \hrule  width \SIZE  height .#3pt
            \vskip -.#3pt
         \repeat } } }

\makeatother
\def\romanno#1{\uppercase\expandafter{\romannumeral#1}}

\begin{document}
\begin{titlepage}
\begin{center}
\begin{Large}
{\bf Determinant Structure of the Rational Solutions}\\[2mm]
{\bf for the Painlev\'e IV Equation}\\[3mm]
\end{Large}
\vspace{30pt}
\begin{normalsize}
{\sc Kenji Kajiwara}\\[3mm]
{\it Department of Electrical Engineering,
Doshisha University,}\\
{\it Tanabe, Kyoto 610-03, Japan}\\[5mm]
{and}\\[5mm]
{\sc Yasuhiro Ohta}\\[3mm]
{\it Department of Applied Mathematics, Faculty of Engineering,}\\
{\it Hiroshima University, }\\
{\it 1-4-1 Kagamiyama, Higashi-Hiroshima 739, Japan}\\
\end{normalsize}
\end{center}
\vspace{30pt}
\begin{abstract}
Rational solutions for the Painlev\'e IV equation
are investigated by Hirota bilinear formalism. It is shown that
the solutions in one hierarchy are expressed by 3-reduced Schur functions,
and those in another two hierarchies by Casorati determinant of the Hermite
polynomials, or by special case of the Schur polynomials.
\end{abstract}

\end{titlepage}

\addtolength{\baselineskip}{.3\baselineskip}

\section{Introduction}
It is known
that the six Painlev\'e equations
P$_{\rm I}$--P$_{\rm VI}$ are the fundamental
equations in the theory of nonlinear integrable systems in wider sense,
and thus their solutions are regarded as the ``nonlinear version 
of special functions''\cite{Book}.
Not only for the use of Painlev\'e equations in physical context,
they also have many interesting mathematical structures,
one of which is the structure of particular solutions.
As for the algebraic solutions, it is known that some of the Painlev\'e equations 
admit rational solutions expressed by classical polynomials, e.g,
Jacobi and Legendre polynomials for P$_{\rm VI}$, Laguerre polynomials
for P$_{\rm V}$ and Hermite polynomials for P$_{\rm IV}$.
However, there also arise some non-classical polynomials, as pointed
out by Umemura\cite{Umemura:pols}.
Typical example is the Yablonskii-Vorobe'v polynomials which appear
in the rational solutions of P$_{\rm II}$,
\begin{equation}
\frac{d^2y}{dt^2} = 2y^3 + ty + \alpha .
\label{P2}
\end{equation}
Let $T_m$ $(m=0,1,\cdots)$ be polynomials generated recursively by
\begin{equation}
\frac{d^2T_m}{dt^2}T_m - \left(\frac{dT_m}{dt}\right)^2=\frac{1}{4}\left(
tT_m^2-T_{m+1}T_{m-1}\right),\quad T_0=T_1=1.
\end{equation}
Then, it is known that 
\begin{equation}
y=\frac{d}{dt}\log \frac{T_{m+1}}{T_m}
\end{equation}
satisfies P$_{\rm II}$ with $\alpha=-m-1$\cite{Yab,Vor}.
The characteristic polynomials $T_m$ are called the Yablonskii-Vorobe'v
polynomials. To clarify the nature of these polynomials,
it is useful to study the relation with the soliton equations.
In fact, P$_{\rm II}$ is derived {}from the similarity reduction of
the modified KdV equation. This fact implies that 
the Yablonskii-Vorobe'v polynomials are closely related to the $\tau$
function which gives the rational solutions of the modified KdV equations.
Based on this expectation, we have shown in the previous paper that
Yablonskii-Vorobe'v polynomials are expressed as the specialization
of the 2-reduced Schur functions\cite{P2_rational}. 

In this article, we investigate the rational
solutions for P$_{\rm IV}$,
\begin{equation}
\frac{d^2w}{dz^2}=\frac{1}{2w}\left(\frac{dw}{dz}\right)^2
+\frac{3}{2}w^3+4zw^2+2(z^2-\alpha)w+\frac{\beta}{w}\ ,
\label{P4}
\end{equation}
where $\alpha$ and $\beta$ are parameters.
There are various simple particular solutions of physical importance\cite{BCH:1},
and it is possible to obtain solutions of ``higher order'' by applying
the B\"acklund transformations\cite{BCH:1,Murata,Fokas,Gromak}
which map one solution to
another solution of P$_{\rm IV}$ with different values of parameters.
As for rational solutions of P$_{\rm IV}$(\ref{P4}), it is known that
there are three hierarchies of unique rational solutions
\cite{Murata}(the name of the hierarchies are due to Ref.\cite{BC}),\\
`` $-{\displaystyle\frac{1}{z}}$ hierarchy'':\\
\begin{equation}
 w=\frac{P_{n-1}(z)}{Q_n(z)}\ ,\quad
(\alpha,\beta)=(\pm k, -2(1+2l+k)^2),\  k,l\in {\bf Z},\ 
l\leq -1,\ k\leq -2l\ ,\label{param1_Murata}
\end{equation}
`` $-2z$ hierarchy'':\\
\begin{equation}
 w=-2z+\frac{P_{n-1}(z)}{Q_n(z)}\ ,\quad
(\alpha,\beta)=(k, -2(1+2l+k)^2),\  k,l\in {\bf Z},\ 
l\geq 0,\ k\leq -l\ ,\label{param2_Murata}
\end{equation}
`` $-{\displaystyle\frac{2}{3}}z$\ hierarchy'':\\
\begin{equation}
 w=-\frac{2}{3}z+\frac{P_{n-1}(z)}{Q_n(z)}\ ,\quad
(\alpha,\beta)=(2k,-2(\pm \frac{1}{3}+2l)^2 ),\  (2k+1,-2(\pm
\frac{2}{3}+2l)^2 ),\ k,l\in {\bf Z},\label{par:okamoto}
\end{equation}
where $P_m(z)$ and $Q_m(z)$ are some polynomials in $z$ of degree $m$,
and that there is no other rational solutions.

Lukashevich has shown that the simplest solutions in the first two
hierarchies are expressed by the Hermite
polynomials\cite{Lukashevich}. Okamoto has studied the B\"acklund transformations
and shown that the log derivative of
ratio of two-directional Wronskians of the Hermite polynomials give the
solutions of P$_{\rm IV}$\cite{Okamoto}.
Moreover, Murata\cite{Murata} has pointed out that any solution in the
$-{\displaystyle\frac{1}{z}}$ hierarchy can be transformed to a solution in the $-2z$
hierarchy, and vice versa. These facts strongly implies that
these two hierarchies have the same nature, and the solutions may be expressed
by log derivative of ratio of some determinants whose entries
are related with the Hermite polynomials. Those determinants are
called the $\tau$ functions.
We investigate these hierarchies by using the Hirota's bilinear formalism and show that
all of the solutions in those hierarchies are expressed by
the $\tau$ functions which is nothing but 
the Casorati determinants (or equivalently, the Wronskians) of
the Hermite polynomials. Moreover, the $\tau$ functions are also expressible
in terms of the Schur functions.

The structure of the solutions in $- \frac{2}{3}z$ hierarchy has been unknown, but
Okamoto\cite{Okamoto} has studied this hierarchy and obtained the following
result.
\begin{prop}[Okamoto]
Let $Q_m,\ m\in{\bf Z}_{\geq 0}$, be polynomials in $x$ generated by 
\begin{equation}
\frac{d^2}{dx^2}\log Q_m + x^2+2m-1=\frac{Q_{m+1}Q_{m-1}}{Q_m^2},\quad
Q_0=Q_1=1.
\end{equation}
Then 
\begin{equation}
u = \frac{d}{dx}\log \frac{Q_{m+1}}{Q_m}-x,
\end{equation}
satisfies P$_{\rm IV}$,
\begin{equation}
\frac{d^2u}{dx^2}=\frac{1}{2u}\left(\frac{du}{dx}\right)^2
+\frac{3}{2}u^3+6xu+\frac{9}{2}\left(x^2-\frac{4}{3}m\right)u
-\frac{1}{2u}\ .
\end{equation}
Similarly, let $R_m,\ m\in{\bf Z}_{\geq 0}$, be polynomials in $x$ generated by 
\begin{equation}
\frac{d^2}{dx^2}\log R_m + x^2+2m=\frac{R_{m+1}R_{m-1}}{R_m^2},\quad
R_0=1,\quad R_1=x.
\end{equation}
Then 
\begin{equation}
u = \frac{d}{dx}\log \frac{R_{m+1}}{R_m}-x,
\end{equation}
satisfies P$_{\rm IV}$,
\begin{equation}
\frac{d^2u}{dx^2}=\frac{1}{2u}\left(\frac{du}{dx}\right)^2+\frac{3}{2}u^3+6xu+\frac{9}{2}\left(x^2-\frac{4}{3}(m+\frac{1}{2})
\right)u-\frac{4}{2u}\ .
 \end{equation}
\end{prop}
The characteristic polynomials $Q_m$ and $R_m$ are called the Okamoto polynomials
\cite{Umemura:pols}. Indeed, the solutions which are expressed by the Okamoto polynomials
are special case of the $-\frac{2}{3}z$ hierarchy.

The key to understand the nature of the Okamoto polynomials lies in the relation
with soliton equations, as is in the case of P$_{\rm II}$. 
In fact, several authors have pointed out the relation between
P$_{\rm IV}$ and the Boussinesq equation\cite{Clarkson:Boussinesq,QNC,Kawamoto,NT}
which belongs to the 3-reduction of the KP hierarchy\cite{JM}.
This implies that some solutions of P$_{\rm IV}$ may be understood as
the similarity reduction of the 3-reduced KP hierarchy. We show that
the solutions in the $-2/3 z$ hierarchy is indeed the case, namely they
are expressed by the 3-reduced Schur functions.

\section{$-{\displaystyle\frac{2}{3}}z$ Hierarchy}
Let us consider the Schur functions in $x_1,x_2,\cdots$ labelled by
Young diagram $Y=(i_1,i_2,i_3,\cdots,i_l)$, $i_1\geq i_2\geq \cdots\geq i_l\geq 0$,
\begin{equation}
S_Y(x_1,x_2,\cdots)=\left\vert
\begin{array}{cccc}
 p_{i_1}&p_{i_1+1} &\cdots &p_{i_1+l-1} \\
 p_{i_2-1}&p_{i_2} &\cdots &p_{i_2+l-2} \\
\vdots & \vdots &\ddots &\vdots\\
 p_{i_l-l+1}&p_{i_l-l+2} &\cdots &p_{i_l} 
\end{array}\right\vert,
\end{equation}
where $p_k$'s are polynomials in $x_1,x_2,\cdots $defined by
\begin{equation}
\sum_{k=0}^\infty p_k(x_1,x_2,\cdots)\lambda^k=\exp\sum_{n=1}^\infty x_k\lambda^k,\quad
p_n=0\ (n<0).
\end{equation}
It is well known that $S_Y$ is a $\tau$ function of the KP hierarchy\cite{JM}.

We consider the 3-reduction of the $\tau$ function, namely, we impose
the condition,
\begin{equation}
\frac{\partial\tau_Y}{\partial x_{3k}}=0,\quad k=1,2,\cdots\ .\label{3-reduction}
\end{equation}
To realize this condition, it is sufficient to restrict the
Young diagram as
\begin{equation}
Y=(M+2n,M+2n-2,\cdots,M+2,M^2,(M-1)^2,\cdots,2^2,1^2), \label{Young1}
\end{equation}
or
\begin{equation}
Y=(N+2n-1,N+2n-3,\cdots,N+1,N^2,(N-1)^2,\cdots,2^2,1^2).\label{Young2}
\end{equation}
In fact, we can easily verify that the Schur functions
associated with the above Young diagrams satisfy the condition
(\ref{3-reduction}), noticing the relation,
\begin{equation}
\frac{\partial p_n}{\partial x_{k}}=p_{n-k}\ .
\end{equation}
For notational simplicity, we rearrange the structure of the Schur polynomials
associated with the Young diagrams (\ref{Young1}) and (\ref{Young2}) as
\begin{equation}
\tau_M^N(x_1,x_2,\cdots)=\left\vert
\begin{array}{ccccccc}
\multicolumn{2}{c}{\dotfill}&p_{2N-1}&p_{2N} &p_{2N+1}& \multicolumn{2}{c}{\dotfill}\\
\multicolumn{3}{c}{\dotfill}&p_{2N-3}&p_{2N-2}& p_{2N-1}&\cdots\\
\ &\ &\ &\ &\ &\hskip-40pt\mbox{$\ddots$} & \ \\
\multicolumn{4}{c}{\dotfill}& p_3      &p_4    & p_5\\
\multicolumn{4}{c}{\dotfill}& p_0      &p_1    & p_2\\
\multispan{7}\hblfil\\
\multicolumn{2}{c}{\dotfill}&p_{2M-2}&p_{2M-1} &p_{2M} &\multicolumn{2}{c}{\dotfill}\\
\multicolumn{3}{c}{\dotfill}&p_{2N-4}&p_{2N-3}& p_{2N-2}&\cdots\\
\ & \ & \ &\  &\  &\hskip-40pt\mbox{$\ddots$} & \ \\
\multicolumn{4}{c}{\dotfill}& p_2      &p_3    & p_4\\
\multicolumn{4}{c}{\dotfill}& p_{-1}   &p_0    & p_1
\end{array}
\right\vert.\label{taumn}
\end{equation}
We can see the equivalence of $\tau_M^N$ and 3-reduced Schur functions
in the following manner,
\begin{equation}
\tau_M^N=
\left\{
\begin{array}{ll}
S_{(M+2n,M+2n-2,\cdots,M+2,M^2,(M-1)^2,\cdots,1^2)} & {\rm for}\ N=M+n>M,\\
S_{(N+2n-1,N+2n-3,\cdots,N+1,N^2,(N-1)^2,\cdots,1^2)}&{\rm for}\ N<M=N+n\ .
\end{array}
\right.
\end{equation}
Then, we have the following theorem:
\begin{th}\label{main0}
Let $x_1=x$, $x_2=\frac{1}{2}$, 
$x_k=0(k=3,4,\cdots)$ in $\tau_M^N$.
Then 
\begin{equation}
 w=\sqrt{\frac{2}{3}}\left\{\left(\log\frac{\tau_{M+1}^N}{\tau_M^N}\right)_x-x\right\},\quad
z=\sqrt{\frac{3}{2}} x ,
\end{equation}
satisfies P$_{\rm IV}$(\ref{P4})
with
\begin{equation}
 (\alpha,\beta)=(2M-N+1,-2\left(N+\frac{2}{3}\right)^2)\quad M,N\in {\bf Z}_{\geq 0}.
\label{par:okamoto1}
\end{equation}
Similarly, 
\begin{equation}
 w=\sqrt{\frac{2}{3}}\left\{\left(\log\frac{\tau_{M}^{N+1}}{\tau_M^N}\right)_x-x\right\},
\quad z=\sqrt{\frac{3}{2}} x,
\end{equation}
satisfies P$_{\rm IV}$ (\ref{P4}) with
\begin{equation}
(\alpha,\beta)=(-M+2N+2,-2\left(M+\frac{1}{3}\right)^2)\quad M,N\in {\bf Z}_{\geq 0}.
\label{par:okamoto2}
\end{equation}
\end{th}
\begin{rem}
The above solutions cover all of the $-\frac{2}{3}z $ hierarchy, which
is easily verified by comparing the parameters (\ref{par:okamoto1}) and (\ref{par:okamoto2})
with (\ref{par:okamoto}).
\end{rem}
Moreover, we have:
\begin{cor}
The Okamoto polynomials are the special case of 
3-reduced Schur functions,
 \begin{equation}
Q_m = \tau_0^m = S_{(2m,2m-2,\cdots,2)}(x,\frac{1}{2},0,\cdots),\quad
R_m = \tau_m^0 = S_{(2n-1,2n-3,\cdots,1)}(x,\frac{1}{2},0,\cdots).
 \end{equation}
\end{cor}
Theorem \ref{main0} is proved by considering the similarity reduction
of 3-reduction of the first modified KP hierarchy, namely, hierarchy of the B\"acklund 
transformations of the KP hierarchy.
First three of the bilinear equations in this
hierarchy are given by
\cite{JM},
\begin{equation}
\left(D_{x_1}^2-D_{x_2}\right)\tau_{M+1}^N\cdot\tau_M^N=0,\label{bl1:3-reduced}
\end{equation}
\begin{equation}
\left(D_{x_1}^3 +3D_{x_1}D_{x_2}\right)\tau_{M+1}^N\cdot\tau_M^N=0,\label{bl2:3-reduced}
\end{equation}
\begin{equation}
\left(D_{x_1}^4-3D_{x_1}^2D_{x_2}-6D_{x_2}^2\right)\tau_{M+1}^N\cdot\tau_M^N=0,
\label{bl3:3-reduced}
\end{equation}
where $D_x$ is the Hirota's bilinear differential operator defined by
\begin{equation}
D_x^n f\cdot g=\left.\left(\partial_x-\partial_{x^\prime}\right)^nf(x)g(x^\prime)
\right|_{x=x^\prime}.
\end{equation}
Next, we apply the similarity reduction. 
\begin{lem}\label{lem1}
Let $\tau_N^M=\tau_N^M(x_1,x_2,0,0,\cdots)$. Then 
 \begin{equation}
\partial_{x_2}\tau_M^N=\frac{1}{2x_2}
\left((M^2-MN+N^2+N)\tau_M^N-x_1\partial_{x_1}\tau_M^N\right).\label{sym}
 \end{equation}
\end{lem}
{\it Proof:} Notice that $p_k$, the entry of $\tau_M^N$,
is a homogeneous polynomial in $(x_1,x_2,x_3,\cdots)$ with degree $k$ if we define
the degree of $x_k$ as $k$. Putting $x_k=0\ (k\geq 3),$
then $\tau_M^N$ is a homogeneous polynomial in $x_1$ and $x_2$ of degree $M^2-MN+N^2+N$.
Thus, if we set
\begin{equation}
f_M^N=\frac{1}{x_2^{(M^2-MN+N^2+N)/2}}\tau_M^N,
\end{equation}
then $f_M^N$ depends only on $t=\frac{x_1}{x_2^{1/2}}$. This implies
\begin{equation}
\partial_{x_2}f_M^N=\frac{\partial t}{\partial x_2}\frac{d}{dt}f_M^N
=-\frac{1}{2}\frac{x_1}{x_2^{3/2}}\frac{d}{dt}f_M^N,
\end{equation}
\begin{equation}
\partial_{x_1}f_M^N=\frac{\partial t}{\partial x_1}\frac{d}{dt}f_M^N
=\frac{1}{x_2^{1/2}}\frac{d}{dt}f_M^N,
\end{equation}
which yields
\begin{equation}
-2x_2\partial_{x_2}f_M^N=x_1\partial_{x_1}f_M^N\ .\label{symf}
\end{equation}
Rewriting eq.(\ref{symf}) in terms of $\tau_M^N$, we get eq. (\ref{sym}).\qed

By using Lemma \ref{lem1}, all the $x_2$ derivatives in 
eqs.(\ref{bl1:3-reduced})-(\ref{bl3:3-reduced}) are rewritten
in terms of $x_1$ derivatives as
\begin{equation}
\left(D_{x_1}^2+\frac{x_1}{2x_2}D_{x_1}-\frac{1}{2x_2}(2M-N+1)\right)
\tau_{M+1}^N\cdot\tau_M^N=0,
\end{equation}
\begin{equation}
\left(D_{x_1}^3-\frac{3}{2x_2}\left(-x_1D_{x_1}^2+(2M-N+1)D_{x_1}+\partial_{x_1}\right)\right)
\tau_{M+1}^N\cdot\tau_M^N=0,
\end{equation}
\begin{eqnarray}
&&\left[D_{x_1}^4-\frac{3}{2x_2}\left(-x_1D_{x_1}^3+(2M-N+1)D_{x_1}^2-2\partial_{x_1}D_{x_1}\right)
\right.\nonumber\\
&&\left. -\frac{6}{4x_2^2}\left(x^2_1D_{x_1}^2-2x_1(2M-N+1)D_{x_1}+3x_1\partial_{x_1}
-(3N^2+4N+1)\right)\right]\tau_{M+1}^N\cdot\tau_M^N=0,
\end{eqnarray}
respectively, where 
\begin{equation}
\partial_{x_1}D_{x_1} f\cdot g=f_{x_1x_1}g-fg_{x_1x_1}\ ,
\end{equation}
\begin{equation}
\partial_{x_1}f\cdot g=\partial_{x_1}(fg)=f_{x_1}g+fg_{x_1}\ .
\end{equation}
Putting $x_1=x$ and $x_2=\frac{1}{2}$, we have:
\begin{prop}
$\tau_M^N=\tau_M^N(x,\frac{1}{2},0,\cdots)$ and
$\tau_{M+1}^N=\tau_{M+1}^N(x,\frac{1}{2},0,\cdots)$ satisfy the following bilinear equations,
\begin{equation}
\left(D_x^2+xD_x-(2M-N+1)\right)
\tau_{M+1}^N\cdot\tau_M^N=0,\label{bl1:sym}
\end{equation}
\begin{equation}
\left(D_x^3-3\left(-xD_x^2+(2M-N+1)D_x+\partial_{x}\right)\right)
\tau_{M+1}^N\cdot\tau_M^N=0,\label{bl2:sym}
\end{equation}
\begin{eqnarray}
&&\left[D_x^4+3xD_x^3-3\{2x^2+(2M-N+1)\}D_x^2
+12x(2M-N+1)D_x\right.\nonumber\\
&&\left. -6\left(-\partial_xD_x+3x\partial_1-(3N^2+4N+1)\right)
\right]\tau_M^N\cdot\tau_{M+1}^N=0.\label{bl3:sym}
\end{eqnarray}
\end{prop}

Now the proof of Theorem \ref{main0} is strait-forward.
Dividing eqs.(\ref{bl1:sym})-(\ref{bl3:sym}) by $\tau_M^N\tau_{M+1}^N$ and using
the formulas\cite{Network},
\begin{eqnarray}
\frac{D_x\tau_M^N\cdot \tau_{M+1}^N}{\tau_M^N\tau_{M+1}^N}&=&\phi_x,\\
\frac{D_x^2\tau_M^N\cdot \tau_{M+1}^N}{\tau_M^N\tau_{M+1}^N}&=&\rho_{xx}+\phi_x^2,\\
\frac{D_x^3\tau_M^N\cdot \tau_{M+1}^N}{\tau_M^N\tau_{M+1}^N}&=&
\phi_{xxx}+3\phi_x\rho_{xx}+\phi_x^3,\\
\frac{D_x^4\tau_M^N\cdot \tau_{M+1}^N}{\tau_M^N\tau_{M+1}^N}&=&
\rho_{xxxx}+4\phi_x\phi_{xxx}+3\rho_{xx}^2+6\phi_x^2\rho_{xx}+\phi_x^4,\\
\frac{\partial_xD_x\tau_M^N\cdot \tau_{M+1}^N}{\tau_M^N\tau_{M+1}^N}&=&
\phi_{xx}+\phi_x\rho_x,
\end{eqnarray}
where $\rho=\log(\tau_M^N\tau_{M+1}^N)$ and $\phi=\log(\tau_{M+1}^N/\tau_M^N)$,
we get
\begin{equation}
\rho_{xx}+\phi_x^2+x\phi_x-(2M-N+1)=0,
\end{equation}
\begin{equation}
\phi_{xxx}+3\phi_x\rho_{xx}+\phi_x^3 - 3x(\rho_{xx}+\phi_x^2)
+3(2M-N+1)\phi_x-3\rho_x=0,
\end{equation}
\begin{eqnarray}
&&\rho_{xxxx}+4\phi_x\phi_{xxx}+3\rho_{xx}^2+6\phi_x^2\rho_{xx}
+\phi_x^4\nonumber\\
&&+3x(\phi_{xxx}+3\phi_x\rho_{xx}+\phi_x^3)
-3(2M-N+1+2x^2)(\rho_{xx}+\phi_x^2)\nonumber\\
&&+12x(2M-N+1)\phi_x+6(\phi_{xx}+\phi_x\rho_x)
-18x\rho_x+6(3N^2+4N+1)=0,
\end{eqnarray}
respectively. Eliminating $\rho$ and putting $u=\phi_x-x$, we obtain P$_{\rm IV}$,
\begin{equation}
\frac{d^2u}{dx^2}=\frac{1}{2u}\left(\frac{du}{dx}\right)^2
+\frac{3}{2}u^3+6xu^2+\left(\frac{9}{2}x^2-3a\right)u
-\frac{3b+1}{2u},\label{p4:okamoto}
\end{equation}
where,
\begin{equation}
a=2M-N+1,\quad b=3N^2+4N+1.
\end{equation}
This gives the half of the Theorem \ref{main0}. Another half is proved in similar
manner by starting {}from the bilinear equations,
\begin{equation}
\left(D_{x_1}^2-D_{x_2}\right)\tau_M^{N+1}\cdot\tau_M^N=0,\label{bl4:3-reduced}
\end{equation}
\begin{equation}
\left(D_{x_1}^3 +3D_{x_1}D_{x_2}\right)\tau_M^{N+1}\cdot\tau_M^N=0,\label{bl5:3-reduced}
\end{equation}
\begin{equation}
\left(D_{x_1}^4-3D_{x_1}^2D_{x_2}-6D_{x_2}^2\right)\tau_M^{N+1}\cdot\tau_M^N=0,
\label{bl6:3-reduced}
\end{equation}
{}from which we get eq.(\ref{p4:okamoto}) for $u=(\log \tau_M^{N+1}/\tau_M^N)-x$
with $a=-M+2N+2$ and $b=3M^2+2M$.
This completes the proof of Theorem \ref{main0}.

 \section{$-{\displaystyle\frac{1}{z}}$ and $-2z$ Hierarchies}
\subsection{$\tau$ Function}
As mentioned in the introduction, solutions in the $-\frac{1}{z}$ and $-2z$ hierarchies
can be transformed each other by B\"acklund transformations. {}From the view of
$\tau$ functions, this fact suggests that their $\tau$ functions are the
same, and only the relation between the $\tau$ functions and the dependent variable
of P$_{\rm IV}$ is different. 
\begin{df}
Let $H_n$ and $\hat H_n$, $n=0,1,2\cdots$, be polynomials in $x$
defined by
\begin{equation}
\sum_{n=0}^\infty \frac{1}{n!}H_n\lambda^n=\exp \left(x\lambda-\frac{1}{2}\lambda^2
\right)\ ,
\end{equation}
\begin{equation}
\sum_{n=0}^\infty \frac{1}{n!}\hat H_n\lambda^n=
\exp \left(x\lambda+\frac{1}{2}\lambda^2\right)\ ,
\end{equation}
respectively. 
Then we define the $\tau$ functions 
$\tau_N^n$ and $\hat\tau_N^n$ to be $N\times N$ determinants given by
\begin{equation}
\tau_N^n=\left\vert
\begin{array}{cccc}
 H_n&H_{n+1} &\cdots &H_{n+N-1} \\
 H_{n+1}&H_{n+2} &\cdots &H_{n+N} \\
 \vdots & \vdots & \ddots &\vdots\\
 H_{n+N-1}&H_{n+N} &\cdots &H_{n+2N-2} 
\end{array}
\right\vert\ ,\label{tau:Hermite}
\end{equation}
 \begin{equation}
\hat \tau_N^n=\left\vert
\begin{array}{cccc}
 \hat H_n&\hat H_{n+1} &\cdots &\hat H_{n+N-1} \\
 \hat H_{n+1}&\hat H_{n+2} &\cdots &\hat H_{n+N} \\
 \vdots & \vdots & \ddots &\vdots\\
 \hat H_{n+N-1}&\hat H_{n+N} &\cdots &\hat H_{n+2N-2} 
\end{array}
\right\vert\ ,\label{tau:m-Hermite}
\end{equation}
respectively.
\end{df}

Then, the solutions in the $-\frac{1}{z}$ hierarchy are
expressed as follows:
\begin{th}\label{main1}
\begin{equation}
w=-\sqrt{2}\left(\log\frac{\tau_{N+1}^n}{\tau_N^n}\right)_x
,\quad z=\frac{1}{\sqrt{2}} x\ ,
\end{equation}
give rational solutions of P$_{\rm IV}$(\ref{P4}) with
\begin{equation}
(\alpha,\beta)=(-(n+2N+1),-2n^2),\quad n,N\in{\bf Z},\quad
n\geq 1,\quad N\geq 0.
\label{param1_kaji}
\end{equation}
Moreover,
\begin{equation}
w=\sqrt{2}\left(\log\frac{\hat\tau_{N+1}^n}{\hat\tau_N^n}\right)_x
,\quad z=\frac{1}{\sqrt{2}} x ,
\end{equation}
give rational solutions of P$_{\rm IV}$(\ref{P4}) with
\begin{equation}
(\alpha,\beta)=(n+2N+1,-2n^2),\quad n,N\in{\bf Z},
\quad n\geq 1,\quad N\geq 0.
\label{param2_kaji}
\end{equation}
\end{th}
Here, several remarks are in order.
\begin{rem}
\begin{enumerate}
\item Parameterization eq.(\ref{param1_kaji}) is equivalent to
eq.(\ref{param1_Murata}) with $\alpha=-k$ if we put $k=n+2N+1$ and
$l=-(N+1)$. Moreover, eq.(\ref{param2_kaji}) is equivalent to
eq.(\ref{param1_Murata}) with $\alpha=k$ if we put $k=n+2N+1$ and
$l=-(N+1)$. 
Hence the solutions in Theorem \ref{main1} cover
all of $-{\displaystyle\frac{1}{z}}$ hierarchy.
\item As shown in the proof, $n$ do not have to be necessarily integers and
hence $H_n$ can be the Hermite-Weber functions.
In fact, we need only the recursion relations,
\begin{equation}
\frac{d}{dx}H_n-xH_n=-H_{n+1},\quad
\frac{d}{dx}H_n=nH_{n-1}.
\label{Hermite}
\end{equation}
\end{enumerate}
\end{rem}
Moreover, $\tau$ function admits several expressions.
\begin{rem}
\begin{enumerate}
\item It is possible to express the solutions in terms of the Wronskian,
since we have the relation,
\begin{eqnarray}
&&\left\vert
\begin{array}{cccc}
 H_n&H_{n+1} &\cdots &H_{n+N-1} \\
 H_{n+1}&H_{n+2} &\cdots &H_{n+N} \\
 \vdots & \vdots & \ddots &\vdots\\
 H_{n+N-1}&H_{n+N} &\cdots &H_{n+2N-2} 
\end{array}
\right\vert\nonumber\\
&=&\prod_{k=1}^{N-1}(n+k)^k 
\left\vert
\begin{array}{cccc}
 H_{n+N-1}&\frac{d}{dx}H_{n+N-1} &\cdots &\left(\frac{d}{dx}\right)^{N-1}H_{n+N-1} \\
 \frac{d}{dx}H_{n+N-1}&\frac{d^2}{dx^2}H_{n+N-1} &\cdots &
\left(\frac{d}{dx}\right)^{N}H_{n+N-1} \\
 \vdots & \vdots & \ddots &\vdots\\
 \left(\frac{d}{dx}\right)^{N-1}H_{n+N-1}&
\left(\frac{d}{dx}\right)^{N}H_{n+N-1} &\cdots &\left(\frac{d}{dx}\right)^{2N-2}H_{n+N-1} 
\end{array}
\right\vert\ 
,\label{Wronskian}
\end{eqnarray}
which can be verified by using eq.(\ref{Hermite}).
\item As is obvious {}from the Wronskian expression eq.(\ref{Wronskian}),  $\tau_N^0$ gives only a constant,
which yields $0$ solution of P$_{\rm IV}$. Thus only the cases of
$n\geq 1$ give nontrivial solutions.
 \item $\tau$ functions are also expressed by the Schur functions. In fact we have, for example,
\begin{equation}
\hat\tau_N^n=\frac{(n+N-1)^!}{(n-1)^!} (-1)^{N(N-1)/2}S_{(n^N)}(x,\frac{1}{2},0,\cdots),
\end{equation}
where,
\[
 k^!=k!^k(k-1)!^{(k-1)}\cdots 2!^21!^1,
\]
which can be verified by noticing
\begin{equation}
\hat H_n=\frac{1}{n!}\ p_n(x,\frac{1}{2},0,\cdots),
\end{equation}
and eq.(\ref{Wronskian}).
\end{enumerate}
\end{rem}

For the solutions in the $-2z$ hierarchy, the following expression
is valid.
\begin{th}\label{main2}
\begin{equation}
w=\sqrt{2}\left\{
\left(\log\frac{\tau_{N+1}^n}{\tau_N^{n+1}}\right)_x
-x\right\} ,\quad z=\frac{1}{\sqrt{2}} x\ ,
\end{equation}
give rational solutions of P$_{\rm IV}$(\ref{P4}) with
\begin{equation}
(\alpha,\beta)=(N-n,-2(n+N+1)^2),\quad n,N\in{\bf Z},\quad
n\geq 0,\quad N\geq 0.
\label{param3_kaji}
\end{equation}
\end{th}
\begin{rem}
\begin{enumerate}
\item Parameterization eq. (\ref{param3_kaji}) is equivalent
to eq.(\ref{param2_Murata}), if we put
$k=N-n$, $l=N$. Hence the solutions in Theorem \ref{main2}
covers all the solutions in the $-2z$ hierarchy.
\item Another $\tau$ function $\hat\tau_N^n$ (\ref{tau:m-Hermite}) can give 
the solution of the same type for P$_{\rm IV}$. In fact, we can show
that
\begin{equation}
\hat w=-\sqrt{2}\left\{
\log\left(\frac{\hat\tau_{N+1}^n}{\hat\tau_N^{n+1}}\right)_x
-x\right\},\quad z= \frac{1}{\sqrt{2}}x\ ,
\end{equation}
satisfies P$_{\rm IV}$ with the parameters $(\alpha,\beta)=
(n-N,-2(n+N+1)^2)$. This parameterization is also equivalent
to eq.(\ref{param2_Murata}), if we put
$k=n - N$, $l=n$. However, the uniqueness of the rational solutions
\cite{Murata,Umemura:P4} implies that they give the same solution as
given in Theorem \ref{main2}. In fact, we can check that
$w$ and $\hat w$ are the same if we exchange $n$ and $N$.
\end{enumerate}
\end{rem}
\subsection{Proof of Theorems}
In this section, we give the proof of the Theorems \ref{main1} and \ref{main2}.
The first half of Theorem \ref{main1} is
a direct consequence of the following proposition.
\begin{prop}\label{prop1}
$\tau_N^n$ satisfies
the following bilinear equations:
\begin{equation}
\left( D_x^2-xD_x+n\right)\tau_{N+1}^n\cdot \tau_N^n=0,
\label{bl1}
\end{equation}
\begin{equation}
\left( D_x^3-6xD_x^2+(5x^2+n-4N-2)D_x+\frac{d}{dx}-5nx\right)
\tau_{N+1}^n\cdot \tau_N^n=0,\label{bl2}
\end{equation}
\begin{eqnarray}
\left( D_x^4-4xD_x^3\right. &+&(11x^2-8n-28N-14)D_x^2-2\frac{d}{dx}D_x
-2x(4x^2-3n-20N-10)D_x\nonumber\\
&-&\left. 3x\frac{d}{dx}+8nx^2-n(9n+28N+14)\right)
\tau_{N+1}^n\cdot \tau_N^n=0\ .\label{bl3}
\end{eqnarray}
\end{prop}
The second half of Theorem \ref{main1} is derived
{}from the following bilinear equations:
\begin{prop}\label{prop2}
$\hat\tau_N^n$ satisfies the following bilinear equations.
 \begin{equation}
\left( D_x^2 + xD_x - n\right)\hat\tau_{N+1}^n\cdot \hat\tau_N^n=0,
\label{mbl1}
 \end{equation}
\begin{equation}
\left( D_x^3 + 6xD_x^2+(5x^2-n+4N+2)D_x-\frac{d}{dx}-5n x\right)
\hat\tau_{N+1}^n\cdot \hat\tau_N^n=0,\label{mbl2}
\end{equation}
\begin{eqnarray}
\left( D_x^4+4xD_x^3\right. &+&(11x^2+8n+28N+14)D_x^2+2\frac{d}{dx}D_x
+2x(4x^2+3n+20N+10)D_x\nonumber\\
&-&\left. 3x\frac{d}{dx}-8nx^2-n(9n+28N+14)\right)
\hat\tau_{N+1}^n\cdot \hat\tau_N^n=0\ .\label{mbl3}
\end{eqnarray}
\end{prop}
Finally, Theorem \ref{main2} is obtained {}from the
following bilinear equations:
\begin{prop}\label{prop3}
$\tau_N^n$ satisfies the
following bilinear equations.
\begin{equation}
\left( D_x^2-xD_x+n-N\right)\tau_{N+1}^n\cdot \tau_N^{n+1}=0,
\label{bl4}
\end{equation}
\begin{equation}
\left( D_x^3-6xD_x^2+(5x^2+n-N)D_x+\frac{d}{dx}-5(n-N)x\right)
\tau_{N+1}^n\cdot \tau_N^{n+1}=0,\label{bl5}
\end{equation}
\begin{eqnarray}
\left( D_x^4\right.&-&4xD_x^3 +(11x^2-8n-28N-18)D_x^2+2\frac{d}{dx}D_x
-2x(4x^2-3n-15N-9)D_x\nonumber\\
&-&\left. x\frac{d}{dx}+8(n-N)x^2-(9n^2+10Nn+14n-27N^2-22N)\right)
\tau_{N+1}^n\cdot \tau_N^{n+1}=0\ .\label{bl6}
\end{eqnarray}
\end{prop}

To prove the Propositions \ref{prop1}-\ref{prop3}, 
we first introduce the notation $\tau_{NY}^n$:
\begin{df}
Let $Y=(i_1,i_2,\cdots,i_h)$ be a Young diagram.
Then we define an $N\times N$ determinant $\tau_{NY}^n$ by
\begin{equation}
\tau_{NY}^n = \left|\matrix{
H_n    &H_{n+1}    &\cdots &H_{n+N-h-1} &H_{n+N-h+i_h}
 &\cdots &H_{n+N-2+i_2} &H_{n+N-1+i_1}\cr
H_{n+1}    &H_{n+2}    &\cdots &H_{n+N-h}   &H_{n+N-h+1+i_h}
 &\cdots &H_{n+N-1+i_2} &H_{n+N+i_1}  \cr
\vdots &\vdots &\cdots &\vdots    &\vdots
 &\cdots &\vdots      &\vdots     \cr
H_{n+N-1}&H_N    &\cdots &H_{n+2N-h-2}&H_{n+2N-h-1+i_h}
 &\cdots &H_{n+2N-3+i_2}&H_{n+2N-2+i_1}}\right|.
\end{equation}
\end{df}
It is possible to derive the bilinear equations
{}from the Pl\"ucker relations which are
identities between the determinants whose columns are shifted.
In fact, Proposition \ref{prop1} is obtained
{}from the following identities:
\begin{lem}\label{lem:pl1}
\begin{equation}
\tau_{N+1}^{n}{}_{\young{11}}\tau_N^n
-\tau^n_{N+1}{}_{\young{1}}\tau^n_N{}_{\young{1}}
+\tau^n_{N+1}\tau^n_N{}_{\young{2}}=0, \label{pl1}
\end{equation}
\begin{equation}
\tau^n_{N+1}{}_{\young{12}}\tau^n_N
-\tau^n_{N+1}{}_{\young{2}}\tau^n_N{}_{\young{1}}
+\tau^n_{N+1}\tau^n_N{}_{\young{3}}=0\ ,\label{pl2}
\end{equation}
\begin{equation}
\tau_{N+1}^{n}\tau_N^n{}_{\young{4}}
-\tau^n_{N+1}{}_{\young{3}}\tau^n_N{}_{\young{1}}
+\tau^n_N\tau^n_{N+1}{}_{\young{13}}=0\ ,\label{pl3}
\end{equation}
\begin{equation}
\tau_{N+1}^{n}{}_{\young{1}}\tau_N^n{}_{\young{3}}
-\tau^n_{N+1}{}_{\young{2}}\tau^n_N{}_{\young{2}}
+\tau^n_N\tau^n_{N+1}{}_{\young{22}}=0\ .\label{pl4}
\end{equation}
\end{lem}
Proposition \ref{prop2} is derived {}from the same identity in $\hat\tau_N^n$.
Similarly, we get Proposition \ref{prop3} {}from the identities,
\begin{lem}\label{lem:pl3}
\begin{equation}
\tau_{N+1}^n{}_{\young{11}}\tau^{n+1}_N
-\tau_{N+1}^n{}_{\young{1}}\tau^{n+1}_N{}_{\young{1}}
+\tau_{N+1}^n\tau^{n+1}_N{}_{\young{2}}=0, \label{pl5}
\end{equation}
\begin{equation}
\tau_{N+1}^n{}_{\young{12}}\tau^{n+1}_N
-\tau_{N+1}^n{}_{\young{2}}\tau^{n+1}_N{}_{\young{1}}
+\tau_{N+1}^n\tau^{n+1}_N{}_{\young{3}}=0\ .\label{pl6}
\end{equation}
\begin{equation}
\tau_{N+1}^{n}\tau_N^{n+1}{}_{\young{4}}
-\tau^n_{N+1}{}_{\young{3}}\tau^{n+1}_N{}_{\young{1}}
+\tau^n_N\tau^{n+1}_{N+1}{}_{\young{13}}=0\ ,\label{pl7}
\end{equation}
\begin{equation}
\tau_{N+1}^{n}{}_{\young{1}}\tau_N^{n+1}{}_{\young{3}}
-\tau^n_{N+1}{}_{\young{2}}\tau^{n+1}_N{}_{\young{2}}
+\tau^{n+1}_N\tau^n_{N+1}{}_{\young{22}}=0\ .\label{pl8}
\end{equation}
\end{lem}
We give the derivations of Lemmas \ref{lem:pl1} and \ref{lem:pl3} in 
the appendix.

We next construct the shift operators which are differential
operators generating $\tau_{NY}^n$ {}from $\tau_N^n$ by using the
technique developed in refs. \cite{OandN} and \cite{P2_rational}. 
In fact, we have:
\begin{lem}\label{le1}
\begin{equation}
\tau_N^n{}_{\young{1}}=\left(-\frac{d}{dx}+xN\right)\tau_N^n.
\label{shift1}
\end{equation}
\end{lem}
{\it Proof.} Notice that $\tau_{N\young{1}}^n$
is expressed by
\begin{equation}
\tau_{N\young{1}}^n = \pmatrix{
H_{n+1} & H_{n+2} &\cdots & H_{n+N}\cr
H_{n+2} & H_{n+3} &\cdots & H_{n+N+1}\cr
\vdots &\vdots &\ddots &\vdots\cr
H_{n+N} & H_{n+N+1} &\cdots & H_{n+2N-1}\cr}\cdot
\pmatrix{
\Delta_{11} & \Delta_{12} &\cdots &\Delta_{1N}\cr
\Delta_{21} & \Delta_{22} &\cdots &\Delta_{2N}\cr
\vdots      & \vdots      &\ddots &\vdots\cr
\Delta_{N1} & \Delta_{N2} &\cdots &\Delta_{NN}\cr}\ ,\label{eq1}
\end{equation}
where $\Delta_{ij}$ is the $(i,j)-$cofactor of $\tau_N^n$ and
$A\cdot B$ denotes a standard scalar product for $N\times N$
matrices $A=(a_{ij})$ and $B=(b_{ij})$ which is defined as
\begin{equation}
A\cdot B=\sum_{i,j=1}^N a_{ij}b_{ij}={\rm Tr}\ A^tB.
\end{equation}
The first matrix of
(\ref{eq1}) is rewritten by using the recursion relation
(\ref{Hermite}) as 
\begin{eqnarray}
&{}&
-\pmatrix{
\partial_x H_{n} &\partial_x H_{n+1} &\cdots &\partial_xH_{n+N-1}\cr
\partial_x H_{n+1} &\partial_x H_{n+2} &\cdots &\partial_xH_{n+N} \cr
\vdots         &\vdots         &\ddots &\vdots           \cr
\partial_x H_{n+N-1} &\partial_x H_{n+N} &\cdots &\partial_xH_{n+2N-2}}
\nonumber\\
&+&
x\pmatrix{
 H_{n} & H_{n+1} &\cdots &H_{n+N-1}\cr
 H_{n+1} & H_{n+2} &\cdots &H_{n+N} \cr
\vdots         &\vdots         &\ddots &\vdots           \cr
 H_{n+N-1} & H_{n+N} &\cdots &H_{n+2N-2}}.
\label{eq2}
\end{eqnarray}
Then applying the dot product to eq.(\ref{eq2}), we get
\begin{equation}
\tau_N^n{}_{\young{1}}=\left(-\frac{d}{dx}+xN\right)\tau_N^n.
\end{equation}
Thus we have proved Lemma \ref{le1}.\qed\\
For the shift operators of second order, we have:
\begin{lem}\label{le2}
\begin{equation}
\tau_N^n{}_{\young{2}}=
\frac{1}{2}\left(\frac{d^2}{dx^2}-(2N+1)x\frac{d}{dx}
+x^2N(N-1) - N(N+n+1)\right)\tau_N^n\ ,
\label{shift2}
\end{equation}
\begin{equation}
\tau_N^n{}_{\young{11}}
=\frac{1}{2}\left(\frac{d^2}{dx^2}-(2N-1)x\frac{d}{dx}
+N(N-1)x^2+N(N+n-1)\right)\tau_N^n\ .
\label{shift3}
\end{equation}
\end{lem}
{\it Proof.} We consider
\begin{eqnarray}
&{}&\tau_{N}^n{}_{\young{2}}+\tau_{N}^n{}_{\young{11}}
=
\pmatrix{
H_{n+1} & H_{n+2} &\cdots & H_{n+N-1}& H_{n+N+1}\cr
H_{n+2} & H_{n+3} &\cdots &H_{n+N}& H_{n+N+2}\cr
\vdots &\vdots &\vdots &\vdots&\vdots\cr
H_{n+N} & H_{n+N+1} &\cdots &H_{n+2N-2}& H_{n+2N}\cr}\nonumber\\
&\cdot&
\pmatrix{
\Delta_{\young{1}}{}_{11} & \Delta_{\young{1}}{}_{12} &\cdots &\Delta_{\young{1}}{}_{1N}\cr
\Delta_{\young{1}}{}_{21} & \Delta_{\young{1}}{}_{22} &\cdots &\Delta_{\young{1}}{}_{2N}\cr
\vdots      & \vdots      &\ddots &\vdots\cr
\Delta_{\young{1}}{}_{N1} & \Delta_{\young{1}}{}_{N2} &\cdots
&\Delta_{\young{1}}{}_{NN}\cr}
\ ,
\end{eqnarray}
where $\Delta_{\young{1}}{}_{ij}$ is $(i,j)$ cofactor of
$\tau_N{}_{\young{1}}$.
The first matrix in the right-hand side is equal to
\begin{eqnarray}
&{}&
-\pmatrix{
\partial_xH_{n} & \partial_xH_{n+1} &\cdots & \partial_xH_{n+N-2}& \partial_xH_{n+N}\cr
\partial_xH_{n+1} & \partial_xH_{n+2} &\cdots &\partial_xH_{n+N-1}& \partial_xH_{n+N+1}\cr
\vdots &\vdots &\vdots &\vdots&\vdots\cr
\partial_xH_{n+N-1} & \partial_xH_{n+N} &\cdots &\partial_xH_{n+2N-1}& \partial_xH_{n+2N-1}\cr}
\nonumber\\
&+&x
\pmatrix{
H_{n} & H_{n+1} &\cdots & H_{n+N-2}& H_{n+N}\cr
H_{n+1} & H_{n+2} &\cdots &H_{n+N-1}& H_{n+N+1}\cr
\vdots &\vdots &\vdots &\vdots&\vdots\cr
H_{n+N-1} & H_{n+N} &\cdots &H_{n+2N-1}& H_{n+2N-1}\cr},
\label{eq3}
\end{eqnarray}
Applying the dot product to eq.(\ref{eq3}), we obtain,
\begin{eqnarray}
\tau_{N}^n{}_{\young{2}}+\tau_{N}^n{}_{\young{11}}
&=&\left(-\frac{d}{dx}+xN\right)\tau_{N\young{1}}^n\nonumber\\
&=&\left(\frac{d^2}{dx^2}-2xN\frac{d}{dx}-N+x^2N^2\right)\tau_N^n\ .
\label{s1}
\end{eqnarray}
Next we consider the following equality,
\begin{equation}
\tau_{N}^n{}_{\young{2}}-\tau_{N}^n{}_{\young{11}}
=\pmatrix{
H_{n+2}&H_{n+3}&\cdots &H_{n+N+1}\cr
H_{n+3}&H_{n+4}&\cdots &H_{n+N+2}\cr
\vdots &\vdots &\ddots &\vdots \cr
H_{n+N+1}&H_{n+N+2}&\cdots &H_{n+2N}}
\cdot
\pmatrix{
\Delta_{11} & \Delta_{12} &\cdots &\Delta_{1N}\cr
\Delta_{21} & \Delta_{22} &\cdots &\Delta_{2N}\cr
\vdots      & \vdots      &\ddots &\vdots\cr
\Delta_{N1} & \Delta_{N2} &\cdots &\Delta_{NN}\cr}.
\label{eqq1}
\end{equation}
The first matrix of the right hand side of (\ref{eqq1}) is
rewritten as
\begin{eqnarray}
&-&
\pmatrix{
\partial_xH_{n+1}    &\partial_xH_{n+2}    &\cdots &\partial_xH_{n+N}\cr
\partial_xH_{n+2}    &\partial_xH_{n+3}    &\cdots &\partial_xH_{n+N+1}\cr
\vdots &\vdots &\ddots &\vdots \cr
\partial_xH_{n+N}&\partial_xH_{n+N+1}&\cdots &\partial_xH_{n+2N-1}}\nonumber\\
&+&x\pmatrix{
H_{n+1}    &H_{n+2}    &\cdots &H_{n+N}\cr
H_{n+2}    &H_{n+3}    &\cdots &H_{n+N+1}\cr
\vdots &\vdots &\ddots &\vdots \cr
H_{n+N}&H_{n+N+1}&\cdots &H_{n+2N-1}}
\label{eqq2}
\end{eqnarray}
The first matrix of eq.(\ref{eqq2}) is rewritten as
\begin{eqnarray}
&{}&
\pmatrix{
(n+1)H_{n}    &(n+2)H_{n+1}    &\cdots &(n+N)H_{n+N-1}\cr
(n+2)H_{n+1}    &(n+3)H_{n+2}    &\cdots &(n+N+1)H_{n+N}\cr
\vdots &\vdots &\ddots &\vdots \cr
(n+N)H_{n+N-1}&(n+N+1)H_{n+N}&\cdots &(n+2N-1)H_{n+2N-2}}
\nonumber\\
&=&
\pmatrix{
(n+1)H_{n}    &(n+1)H_{n+1}    &\cdots &(n+1)H_{n+N-1}\cr
(n+2)H_{n+1}    &(n+2)H_{n+2}    &\cdots &(n+2)H_{n+N}\cr
\vdots &\vdots &\ddots &\vdots \cr
(n+N)H_{n+N-1}&(n+N)H_{n+N}&\cdots &(n+N)H_{n+2N-2}}
\nonumber\\
&+&
\pmatrix{
0   &H_{n+1}    &\cdots &(N-1)H_{n+N-1}\cr
0   &H_{n+2}    &\cdots &(N-1)H_{n+N}\cr
\vdots &\vdots &\ddots &\vdots \cr
0   &H_{n+N}    &\cdots &(N-1)H_{n+2N-2}}
\label{eqq3}
\end{eqnarray}
Applying the dot product to eq.(\ref{eqq3}), we get
\[
\left[\left\{ (n+1)+\cdots +(n+N)\right\}+
1+\cdots+(N-1)\right]\tau_N^n
=N(N+n)\tau_N^n\ .
\]
Moreover, {}from the second term of eq.(\ref{eqq2}), we have
\[
x\tau_{N\young{1}}=x\left(-\frac{d}{dx}+xN\right)\tau_N^n\ .
\]
Finally, we obtain
\begin{equation}
\tau_{N}^n{}_{\young{2}}-\tau_{N}^n{}_{\young{11}}
=\left(-x\frac{d}{dx}+x^2N-N(N+n)\right)\tau_N^n\ .\label{s2}
\end{equation}
{}From eqs.(\ref{s1}) and (\ref{s2}), we obtain eqs.
(\ref{shift2}) and (\ref{shift3}).
Thus we have proved Lemma \ref{le2}.\qed\\

Continuing the similar but tedious calculations, we get the following
shift operators of third order:
\begin{lem}\label{le3}
\begin{eqnarray}
\tau^n_N{}_{\young{3}}
&=&\frac{1}{6}\left[ - \frac{d^3}{dx^3}  + 3(N+1)x\frac{d^2}{dx^2}
    + ( -(3N^2+6N+2)x^2 + 3N^2 + 3n N + 7N + 2n + 3)\frac{d}{dx}\right.\nonumber\\
&+& \left.Nx( (N+1)(N+2)x^2 - 3n N - 3N^2 - 4n - 9N -
6)\right]\tau_N^n\ ,
\label{shift4}\\
\tau^n_N{}_{\young{12}}
&=&\frac{1}{3}\left[-\frac{d^3}{dx^3}
           + 3Nx\frac{d^2}{dx^2}
           + ( (-3N^2 + 1)x^2 - n + N )\frac{d}{dx}
           + Nx((N^2-1)x^2 + 2n  )\right]\tau_N^n\ ,
\label{shift5}\\
\tau^n_N{}_{\young{111}}
&=& \frac{1}{6}\left[- \frac{d^3}{dx^3}
  + 3x(N - 1)\frac{d^2}{dx^2}
  + ( (- 3N^2 + 6N - 2)x^2 
       - 3n N + 2n - 3N^2 + 7N - 3)\frac{d}{dx}\right.\nonumber\\
&+&\left. Nx( ( N^2- 3N + 2)x^2 + 3n N - 4n + 3N^2 - 9N +
6)\right]\tau_N^n\ .
\label{shift6}
\end{eqnarray}
\end{lem}

Now we are ready to prove the Proposition \ref{prop1}.
Substituting the shift operators in Lemmas
\ref{le1} and \ref{le2} into the identities (\ref{pl1}) and (\ref{pl2}) in Lemma \ref{lem:pl1},
we get eqs. (\ref{bl1}) and (\ref{bl2}).  
Equation (\ref{bl3}) is obtained in similar manner by using the shift operators
of fourth order, which will be given in the appendix. This completes the proof 
of Proposition \ref{prop1}. Proposition \ref{prop3} is proved {}from the identities in 
Lemma \ref{lem:pl3}. Finally,  to prove Proposition \ref{prop2}, it is necessary to calculate
the shift operators for $\hat\tau_N^n$. 
We omit the detailes, but this is done just by replacing the recursion
relations eq.(\ref{Hermite}) by
\begin{equation}
\frac{d}{dx}\hat H_n+x\hat H_n=\hat H_{n+1},\quad
\frac{d}{dx}\hat H_n=n\hat H_{n-1}.
\label{hat_Hermite}
\end{equation}
Thus proof of Theorems \ref{main1} and \ref{main2} is completed.
\section{Concluding Remarks}
In this article, we have investigated the hierarchies of
rational solutions for P$_{\rm IV}$, and shown that
\begin{enumerate}
 \item Solutions in $-2/3 z$ hierarchy are expressed in terms of
3-reduced Schur functions. In particular, the Okamoto polynomials
are nothing but their special cases.
 \item Solutions in $-2z$ and $-1/z$ hierarchies are expressed in
terms of Casorati determinant of the Hermite polynomials. Moreover,
they are also expressed by special cases of the Schur functions.
\end{enumerate}
It might be an important and interesting problem to characterize the non-classical
polynomials which appears in the algebraic solutions of the Painlev\'e equations
listed in \cite{Umemura:pols} by studying the determinant expressions with
the aid of results of the soliton theory.

We finally note that after obtaining the results, the authors were informed 
that Noumi and Yamada have independently obtained the Schur function expression
for the rational solutions of P$_{\rm IV}$\cite{Noumi}.
\section*{Acknowledgement}
The authors are grateful to Professors  B. Grammaticos, A. Ramani and
J. Hietarinta for discussions and encouragement. 
They also thank Professors H. Umemura and K. Okamoto for 
their interests in this work and discussions.
One of the authors(K.K) was
supported by the Grant-in-aid for Encouragement of Young Scientist
, The Ministry of Education, Science, Sports and Culture of Japan, No. 09740164.
\appendix
\section{Derivation of Pl\"ucker Relations}
We consider the following identity of $(2N+2)\times (2N+2)$
determinant,
\begin{equation}
0=\left\vert\matrix{
 \matrix{0 &1 &\cdots &N-1} &\vbl{4} &\matrix{\hbox{\O}}
&\vbl{4} &N &N+1 &\phi_1 \cr
 \multispan{7}\hblfil \cr
 \matrix{\hbox{\O}} &\vbl{4} &\matrix{0 & \cdots & N-2}
&\vbl{4} &N&N+1 &\phi_1
}\right\vert\ ,\label{id1}
\end{equation}
where ``$k$'' denotes the column vector,  
\begin{equation}
``k"= \left(\matrix{ H_{n+k}\cr H_{n+k+1}\cr \vdots \cr H_{n+k+N-1}}\right),
\end{equation}
and
\begin{equation}
\phi_1=\left(\matrix{ 0\cr 0\cr \vdots \cr 1}\right)\ .
\end{equation}
Applying the Laplace expansion on the right hand side of
eq.(\ref{id1}), we get
\begin{eqnarray}
0&=&|0,\cdots,N-2,N-1,N|\times|0,\cdots N-2,N+1,\phi_1|\nonumber\\
&-&|0,\cdots,N-2,N-1,N+1|\times|0,\cdots N-2,N,\phi_1|\nonumber\\
&+&|0,\cdots,N-2,N-1,\phi_1|\times|0,\cdots N-2,N,N+1|\nonumber\\
&=& 
\tau^n_{N+1}\tau^n_N{}_{\young{2}}
-\tau^n_{N+1}{}_{\young{1}}\tau^n_N{}_{\young{1}}
+\tau_N^n\tau_{N+1}^{n}{}_{\young{11}}
\end{eqnarray}
which is nothing but eq.(\ref{pl1}). Similarly, eq.(\ref{pl2})
is derived {}from the following identity,
\begin{equation}
0=\left\vert\matrix{
 \matrix{0 &1 &\cdots &N-1} &\vbl{4} &\matrix{\hbox{\O}}
&\vbl{4} &N &N+2 &\phi_1 \cr
 \multispan{7}\hblfil \cr
 \matrix{\hbox{\O}} &\vbl{4} &\matrix{0 & \cdots & N-2}
&\vbl{4} &N&N+2 &\phi_1
}\right\vert\ .\label{id2}
\end{equation}
Moreover, we have the following higher order identities:
\begin{eqnarray}
0&=&\left\vert\matrix{
 \matrix{0 &1 &\cdots &N-1} &\vbl{4} &\matrix{\hbox{\O}}
&\vbl{4} &N &N+3 &\phi_1 \cr
 \multispan{7}\hblfil \cr
 \matrix{\hbox{\O}} &\vbl{4} &\matrix{0 & \cdots & N-2}
&\vbl{4} &N&N+3 &\phi_1
}\right\vert\ \nonumber\\
&=&|0,\cdots,N-2,N-1,N|\times|0,\cdots N-2,N+3,\phi_1|\nonumber\\
&-&|0,\cdots,N-2,N-1,N+3|\times|0,\cdots N-2,N,\phi_1|\nonumber\\
&+&|0,\cdots,N-2,N-1,\phi_1|\times|0,\cdots N-2,N,N+3|\nonumber\\
&=& 
\tau_{N+1}^{n}\tau_N^n{}_{\young{4}}
-\tau^n_{N+1}{}_{\young{3}}\tau^n_N{}_{\young{1}}
+\tau^n_N\tau^n_{N+1}{}_{\young{13}},\label{id3}
\end{eqnarray}
\begin{eqnarray}
0&=&\left\vert\matrix{
 \matrix{0 &1 &\cdots &N-1} &\vbl{4} &\matrix{\hbox{\O}}
&\vbl{4} &N+1 &N+2 &\phi_1 \cr
 \multispan{7}\hblfil \cr
 \matrix{\hbox{\O}} &\vbl{4} &\matrix{0 & \cdots & N-2}
&\vbl{4} &N+1&N+2 &\phi_1
}\right\vert\ \nonumber\\
&=&|0,\cdots,N-2,N-1,N+1|\times|0,\cdots N-2,N+2,\phi_1|\nonumber\\
&-&|0,\cdots,N-2,N-1,N+2|\times|0,\cdots N-2,N+1,\phi_1|\nonumber\\
&+&|0,\cdots,N-2,N-1,\phi_1|\times|0,\cdots N-2,N+1,N+2|\nonumber\\
&=& 
\tau_{N+1}^{n}{}_{\young{1}}\tau_N^n{}_{\young{3}}
-\tau^n_{N+1}{}_{\young{2}}\tau^n_N{}_{\young{2}}
+\tau^n_N\tau^n_{N+1}{}_{\young{22}}\ .\label{id4}
\end{eqnarray}

Identities between $\tau_{N+1 Y}^n$ and $\tau_{NY}^{n+1}$ are
derived only by replacing $\phi_1$ in the above identities by
\begin{equation}
\phi_2=\left(\matrix{ 1\cr 0\cr \vdots \cr 0}\right)\ .
\end{equation}
\section{List of the Shift Operators}
\begin{eqnarray}
\tau^n_N{}_{\young{4}}
&=&\frac{1}{24}\left[
\frac{d^4}{dx^4}
- 2(2N + 3)x\frac{d^3}{dx^3}
+ \left\{ (6N^2 + 18N + 11 )x^2
           - 6n N - 8n  - 6N^2 - 22N - 18\right\}\frac{d^2}{dx^2}
\right.\nonumber\\
&+& x\left\{ (- 4N^3 - 18N^2 - 22N - 6  )x^2
+ 12N^3 +12n N^2 + 58 N^2 + 30n N + 12 n + 78N +
25\right\}\frac{d}{dx}\nonumber\\
&+&N\left\{ ( N^3+ 6N^2+ 11N + 6 )x^4
- (6n N^2 + 22n N+ 18n + 6N^3+ 36N^2 + 66N + 36   )x^2\right.\nonumber\\
&+&\left.  \left. 3N^3 + 3n^2N + 6n N^2 + 18N^2 + 6n^2 
+ 24n N + 22n + 33N + 18\right\}\right]\tau_N^n\ ,
\end{eqnarray}
\begin{eqnarray}
\tau^n_N{}_{\young{13}}
&=&\frac{1}{8}\left[
\frac{d^4}{dx^4}
- 2x( 2N + 1)\frac{d^3}{dx^3}
+ \left\{ (6N^2 + 6N - 1 )x^2 - 2n N  - 2N^2 - 6N - 2\right\}
 \frac{d^2}{dx^2}\right.\nonumber\\
&+& x\left\{ (- 4N^3 - 6N^2 + 2N + 2 ) x^2 
+ 4n N^2 - 2n N - 4n + 4N^3 + 10N^2
 + 2N - 3\right\} \frac{d}{dx}\nonumber\\
&+&N\left\{ ( N^3 + 2N^2 - N - 2 ) x^4
+ (- 2N^3 - 4N^2 - 2n N^2 + 2n N + 6n + 2N + 4  ) x^2\right.\nonumber\\
&-& \left.\left.  N^3 + n^2N - 2n N^2 - 2N^2 - 2n^2 - 4n N - 2n 
+ N + 2\right\}\right]\tau_N^n\ , \\
\tau^n_N{}_{\young{22}}
&=&\frac{1}{12}\left[
\frac{d^4}{dx^4}
 - 4Nx \frac{d^3}{dx^3}
+ \left\{ (6N^2 - 1 ) x^2 + 4n + 2N \right\}\frac{d^2}{dx^2}
\right.\nonumber\\
&+&  x\left\{ 2N(- 2N^2 + 1 ) x^2  - 6n N - 2N^2 +
1\right\}\frac{d}{dx}
\nonumber\\
&+& \left.N\left\{ N(N^2 - 1) x^4 + 2n Nx^2
+ 3N^3 + 3n^2N + 6n N^2 - 2n - 3N\right\}\right]\tau_N^n\ ,
\end{eqnarray}
\begin{eqnarray}
\tau^n_N{}_{\young{112}}
&=&\frac{1}{8}
\left[
\frac{d^4}{dx^4} 
+ 2x( - 2N + 1)\frac{d^3}{dx^3} 
+ \left\{ ( 6N^2 - 6N - 1)x^2 + 2n N + 2N^2 - 6N + 2\right\} \frac{d^2}{dx^2}
\right.\nonumber\\
&+& x\left\{ (- 4N^3 + 6N^2 + 2N - 2  )x^2
- 4N^3  - 4n N^2 + 10N^2 - 2n N + 4n - 2N -
3\right\}\frac{d}{dx}\nonumber\\
&+&\left\{ ( N^3 - 2N^2 - N + 2  )x^4 
+ (2n N^2 + 2n N - 6n  + 2N^3 - 4N^2 - 2N + 4 )x^2\right.\nonumber\\
&-&\left.\left. n^2N + 2n^2 - 2n N^2 + 4n N 
- 2n - N^3 + 2N^2 + N - 2\right\}\right]\tau_N^n\ ,\\
\tau^n_N{}_{\young{1111}}
&=&
\frac{1}{24}\left[
\frac{d^4}{dx^4} + 2x( - 2N + 3)\frac{d^3}{dx^3}
+ 
\left\{ (6N^2- 18N + 11)x^2 + 6N^2 + 6n N - 22N - 8n + 18
\right\}\frac{d^2}{dx^2}\right.\nonumber\\
&+& x\left\{ (- 4N^3 + 18N^2 - 22N+ 6 )x^2 
- 12N^3 - 12n N^2 + 58N^2 + 30n N - 12n - 78N + 25\right\}\frac{d}{dx}
\nonumber\\
&+&N\left\{ (N^3 - 6N^2 + 11N - 6) x^4
+ (6N^3 + 6n N^2 - 36N^2 - 22n N + 18n + 66N - 36
)x^2 \right.\nonumber\\
&+&\left.\left. 3N^3 +3n^2N + 6n N^2 - 18N^2 - 6n^2 - 24n N + 22n  + 33N -
18\right\}\right]\tau_N^n.
\end{eqnarray}

\end{document}